\documentclass[sigconf]{acmart}
\usepackage{subcaption}
\usepackage{grffile}
\usepackage{amsmath} 
\usepackage{amssymb}  
\usepackage{xcolor}
\usepackage{booktabs}
\usepackage{algorithm}
\usepackage{algorithmic}

\AtBeginDocument{%
  \providecommand\BibTeX{{%
    \normalfont B\kern-0.5em{\scshape i\kern-0.25em b}\kern-0.8em\TeX}}}

\setcopyright{rightsretained}
\copyrightyear{2020}
\acmYear{2020}
\acmDOI{}

\acmConference[KDD 2020 Workshop on Humanitarian Mapping]{Humanitarian Mapping: Harnessing Data and Human-Machine Intelligence for Actionable Policy Decisions}{August 24, 2020}{San Diego, CA}
\acmBooktitle{KDD 2020 Workshop on Humanitarian Mapping, August 24, 2020, San Diego, CA}
\acmPrice{}
\acmISBN{}

\begin{document}

\title{Fair navigation planning: a humanitarian robot use case}

\author{Martim Brand\~ao}
\email{martim.brandao@kcl.ac.uk}
\orcid{orcid}
\affiliation{%
  \institution{King's College London}
  \city{London}
  \state{UK}
}

\renewcommand{\shortauthors}{Brand\~ao}

\begin{abstract}
In this paper we investigate potential issues of fairness related to the motion of mobile robots. We focus on the particular use case of humanitarian mapping and disaster response.
We start by showing that there is a fairness dimension to robot navigation, and use a walkthrough example to bring out design choices and issues that arise during the development of a fair system. 
We discuss indirect discrimination, fairness-efficiency trade-offs, the existence of counter-productive fairness definitions, privacy and other issues. Finally, we conclude with a discussion of the potential of our methodology as a concrete responsible innovation tool for eliciting ethical issues in the design of autonomous systems.
\end{abstract}




\maketitle


\section{Introduction}
\label{sec:intro}

In recent years there has been a proliferation of research concerned with the ethics of autonomous systems and artificial intelligence \cite{Holzapfel2018,Golub2013,Lin2015}. 
For example, recent studies have shown that machine learning algorithms in recidivism prediction \cite{Angwin2016,Chouldechova2017}, gender classification \cite{Buolamwini2018}, and other applications \cite{Brandao2019fatecv} can perform better for some people compared to others.
This concern has led to greater pressure on developers to innovate responsibly, as well as to the development of a great number of guidelines and principles for ethical development \cite{Winfield2018}. Such guidelines and principles are helpful in providing a framework for researchers and practitioners. However, they are limited in terms of supporting to implement and satisfy the principles in practice. For example, it is often not clear how ``fairness'' or ``beneficence'' principles are relevant to a new technology in practice.

Examples of social inequalities produced by seemingly fairness-unrelated decision-making are numerous and extend well past recent machine learning developments. In the field of ``environmental justice'' \cite{Walker2012}, for example, researchers have argued that the implementation of some transportation policies, which could supposedly improve mobility and access to jobs, can indirectly reinforce inequalities of opportunities \cite{Golub2013}. Other work has shown that waste management sites are often concentrated on low-income, and high racial-minority-percentage locations \cite{Walker2012}. Often such policies do no overtly target such populations, and inequalities of access or exposure to harm can happen because people are not uniformly distributed across space and are in fact usually distributed in ways that relate to economic, cultural and racial factors \cite{Golub2013,Walker2012}. Discrimination is also often embedded within housing markets and the organization of institutions \cite{Walker2012}, which can implicitly influence decision-making and decision outcomes.

These particular discussions of spatially-organized inequalities also relate strongly to mobile robotics. 
Similar issues of environmental discrimination can be found, as we will see, in reconnaissance robots, search-and-rescue robots and humanitarian robots, as they could provide different benefits for high- and low-populated areas, gentrified and young areas, etc. 

In this paper, which summarizes a previous publication \cite{Brandao2020aij}, we investigate the concept of fairness in a humanitarian mapping robot use case.
We use this example to make several claims about fair design in robot navigation, such as:
\begin{enumerate}
	\item Robot navigation paths can give rise to concerns of distributive fairness and indirect-discrimination;
	\item Fairness-aware navigation planners will involve efficiency-fairness trade-offs;
	\item Enforcing certain formalizations of fairness in path planning can be counter-productive;
	\item Design should be an iterative process of understanding the fairness issues of the context at hand;
	\item Mathematical formalization and technology-deployment simulations can be a good tool to draw out fairness-related design decisions with stakeholders in the early stages of design.
\end{enumerate}


\section{Fairness in robot navigation}
\label{sec:navigation}


\subsection{Walkthrough mapping-robot example}
\label{sec:navigationExample}

Imagine a robot that is deployed in the aftermath of a humanitarian disaster in order to find victims that need to receive support, and communicate their location to a response team. This could be a drone searching for earthquake victims. Let us call this a ``rescue drone''.
The drone departs from a base station and needs to go back to the same station for re-charging batteries after a maximum distance is covered. The drone does several of these trips, although we focus on a single trip in isolation. 

\begin{figure}[t]
	\centering
	\def\tc{0} 
	\def\bc{0} 
	\def\lc{2.5} 
	\def\rc{2.5} 
	\begin{subfigure}[b]{0.32\columnwidth}
		\includegraphics[width=\columnwidth, trim=\lc cm \bc cm \rc cm \tc cm, clip=true]{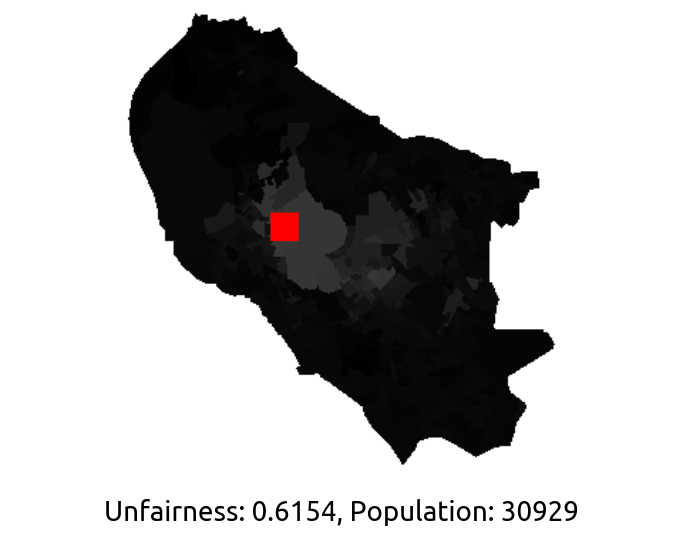}\\
		\includegraphics[width=\columnwidth]{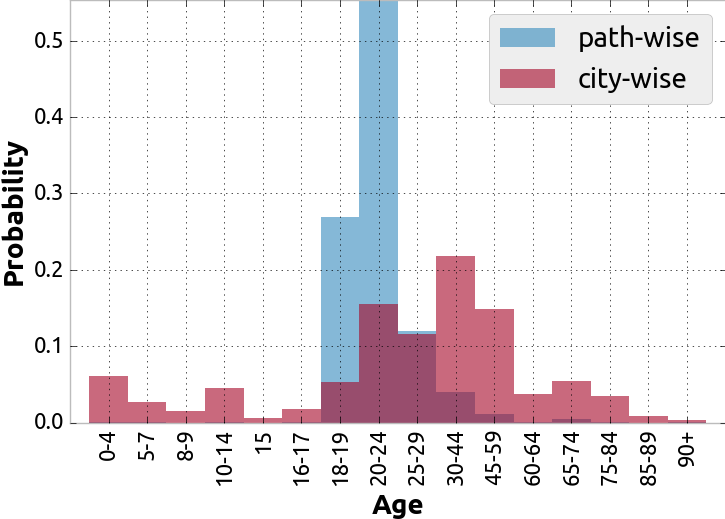}
		\caption{Age}
	\end{subfigure}
	~ 
	\begin{subfigure}[b]{0.32\columnwidth}
		\includegraphics[width=\columnwidth, trim=\lc cm \bc cm \rc cm \tc cm, clip=true]{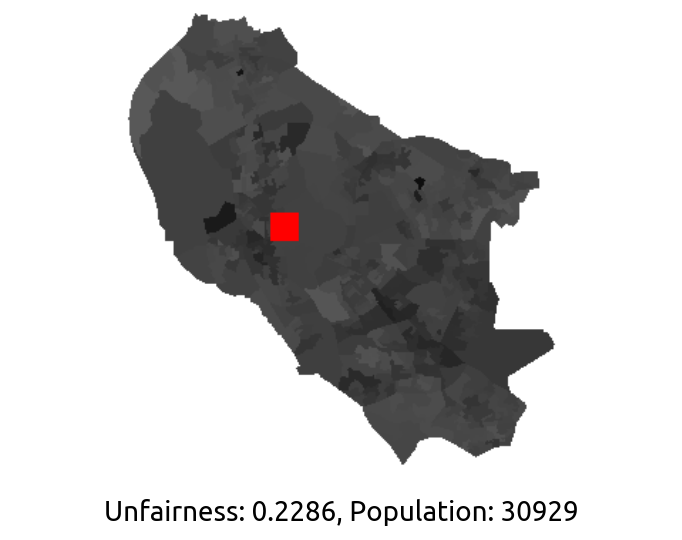}\\
		\includegraphics[width=\columnwidth]{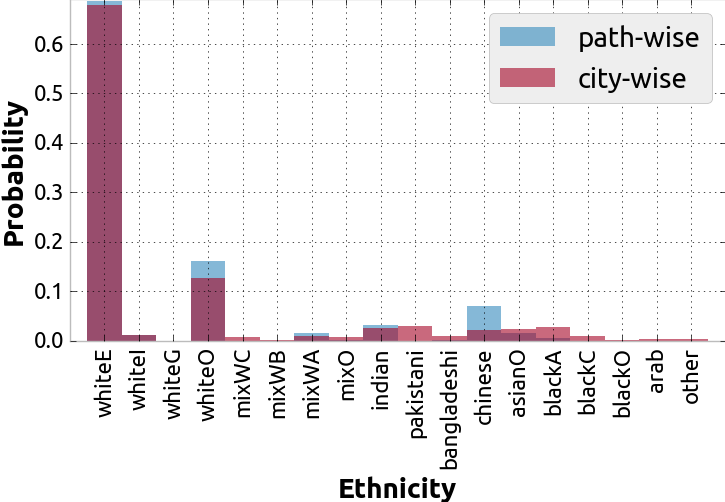}
		\caption{Ethnicity}
	\end{subfigure}
	~
	\begin{subfigure}[b]{0.32\columnwidth}
		\includegraphics[width=\columnwidth, trim=\lc cm \bc cm \rc cm \tc cm, clip=true]{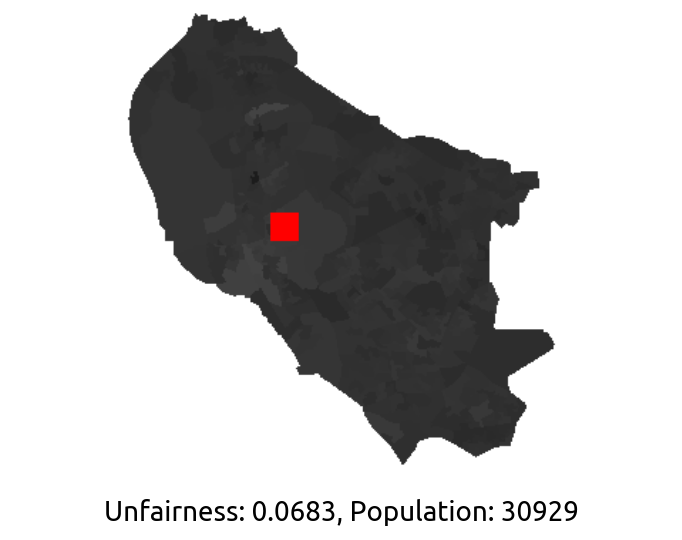}\\
		\includegraphics[width=\columnwidth]{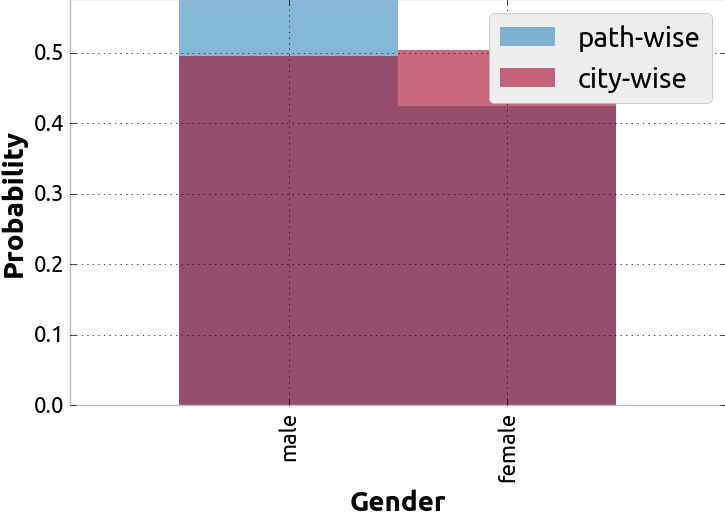}
		\caption{Gender}
	\end{subfigure}
	\caption{Distribution of age, ethnicity and gender over the whole city and a robot path around the city center.}
	\label{fig:fairnessdimension}
\end{figure}

Let us say this happens in a specific location---the city of Oxford, UK---and we use census data to guide the robot towards high population-density areas to increase efficiency. Population density is not uniformly distributed in space. Furthermore, the region has spatial biases related to age, ethnicity and gender, as shown in Figure~\ref{fig:fairnessdimension}. Like other cities \cite{Golub2013}, Oxford has neighborhoods of higher concentration of minority ethnicities and of older populations than the city center.
The consequence for our rescue robot example is the following: planned paths will have skewed distributions of these personal characteristics. If for example the drone thoroughly explores the area immediately around the base station it will find many people because of population density, however most of whom are young and healthy.

Let us suppose that the navigation planner is such that it thoroughly explores the region around the base-station because it is highly populated. 
Figure~\ref{fig:fairnessdimension} shows this path on top of the city map, as well as the path-wise and city-wise personal characteristic distributions of age, ethnicity and gender.
Figure~\ref{fig:fairnessdimension} shows that, as qualitatively seen on the maps, the distribution of age in the center (along the robot's path) is highly biased towards that of undergraduate students, while the city-wide distribution is considerably more uniform. 
The figure also shows that both the center and the city as a whole are highly biased towards a ``white English'' ethnicity. The center has an overrepresented ``white other'' and ``Chinese'' population compared to the rest of the city.


\subsection{Issues raised by this example}
\label{sec:navigationIssues}

\subsubsection{Indirect discrimination within robot navigation}

The example shows that robot navigation paths can be biased in favor or disfavor of different people, as paths inherit spatial distribution biases. 
In particular, the probability of being found by the robot was strongly correlated with age and ethnicity.

In disaster response, such a robot could continue or even reinforce common criticisms in disaster response missions: that policies for selecting disaster response locations usually benefit particular groups of people \cite{DisasterBioethics2013}. Avoiding such discrimination explicitly through algorithms could be a way not only to promote distributive justice, but also to enforce a certain degree of political or commercial neutrality in disaster response (i.e. to make sure that disaster response agencies using robots do not favor a particular group).

\subsubsection{Inequality can be unfair}

While one of the goals of a humanitarian rescue robot is to find as many people as possible, notions of priority also exist \cite{DisasterBioethics2013}. One accepted principle is to attend to the people most-at-risk first \cite{Merin2010,DisasterBioethics2013}.
These could be people with the lowest health-state, those living in low-quality accommodation susceptible of collapse, those that have lower chances of survival if not attended to, such as children and older adults, etc.
In our particular example, the fact that there is indirect discrimination of age, with a bias towards the younger population, is unfair according to such view of disaster response ethics.
So while our robot example is doing part of a disaster response team's job---finding as many people as possible---it is not respecting the context's notion of distributive fairness.

This example also raises the issue of identifying in which personal characteristics indirect discrimination is unfair.
What about the case where a rescue robot finds many people, primarily those at highest risk, but at the same time is biased towards the white population of the city---since it is slightly closer to the base station than the neighborhoods of high minority-concentration?
Should this give rise to concern?
According to defendants of affirmative action it should, since it reinforces social inequalities that have been ingrained in society (through urban policy, housing markets, etc.) for generations.

It then becomes important to provide stakeholders with tools to help identify these possible inequality outcomes of robot deployment so that better design decisions can be made before robot deployment.
One such tool is the methodology that we use in this paper---of simulating the robot's deployment and predicting possible inequalities that can give rise to concerns of distributive fairness.
The example further raises the issue, however, of how one will reach the ``final'' choice of fairness principles and protected characteristics to use in an algorithm. This requires involvement of all stakeholders in the decision process.

\subsubsection{Robots must face dilemmas that humans already face}

Thinking about the fairness issues related to disaster response locations is not something new about rescue robots, but an inherent concern of disaster response itself.
For example, discussions and claims of unfairness are raised regarding disaster response hospital locations when they favor people of specific backgrounds, sometimes politically or commercially favorable to the country backing the response \cite{DisasterBioethics2013}. 
Principles such as most-at-risk first \cite{Merin2010,DisasterBioethics2013} are already used in the field by human doctors.
Thus, the rescue robot example also shows that when robots are used to solve problems currently solved by humans, they must face similar dilemmas currently faced by human decision-makers.


\section{Designing fair robot navigation}
\label{sec:designing}


\subsection{Objects of fairness in navigation}
\label{sec:designingObjects}

We now briefly formalize fairness in the context of robot navigation.
We distinguish between different objects of fairness in navigation: 
\\
\textbf{Locations:} Being fair to locations (e.g. neighborhoods) means that we care about which locations will be visited by a robot: in particular, how often they will be visited, either in absolute value or in comparison to other locations.
\\
\textbf{Protected characteristics:} Being fair to protected characteristics means that we care about the distribution of protected characteristics of the people found along the robot's path.


\subsection{Specifications of fairness in navigation}
\label{sec:designingSpecifications}

In \cite{Brandao2020aij} we provide formal specifications for a set of distributive justice principles in the context of navigation. These are, for example:
\\
\textbf{Demographic parity:} Enforcing (or minimizing the distance to) demographic parity. In the context of navigation this implies that the event of a person being found along a robot's path is independent from group membership (i.e. a protected characteristic).
\\	
\textbf{Rawlsian egalitarian fairness:} maximizing the utility of the worst-off, i.e. maximizing the visit counts of the least-visited region, or the probability of being found for the least-likely group. 
\\	
\textbf{Affirmative action:} Enforcing a desired distribution of location-visits (e.g. region of priority 1 visited M times more than region of priority 2) or protected characteristics (e.g. ratio of younger and older people found along the planned path). This definition is general enough to encompass demographic parity---which is equivalent to a preference towards the distribution of the whole population.


\subsection{Developing a fair navigation planner} 
\label{sec:designingDeveloping}

In \cite{Brandao2020aij} we reach several conclusions regarding the issues involved in the design of a fair system.
Here is a brief summary:

\subsubsection{Fairness may be infeasible, requires trade-offs}

\begin{figure}[t]
	\centering
	\def\tc{0.5} 
	\def\bc{0.5} 
	\def\lc{0.5} 
	\def\rc{0.5} 
	\includegraphics[trim=\lc cm \bc cm \rc cm \tc cm, clip=true, width=0.65\columnwidth]{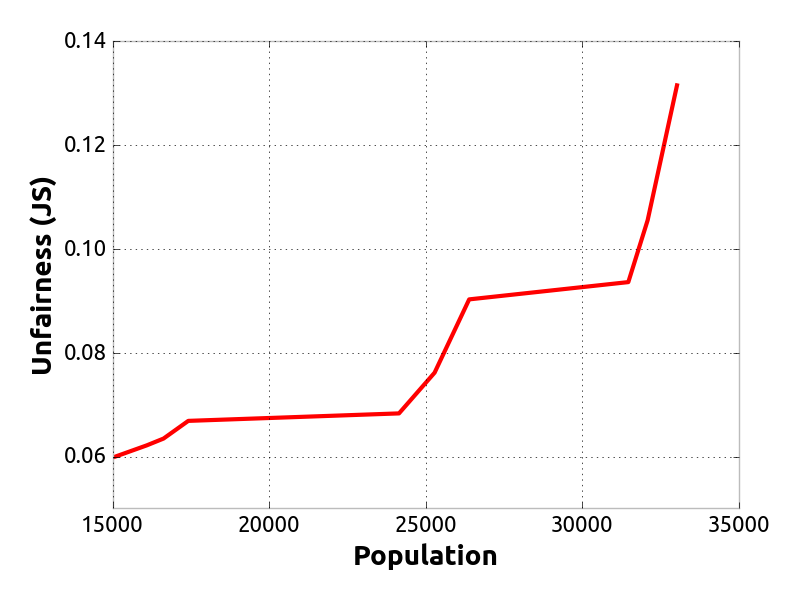}
	\caption{Pareto-curve showing the trade-off between the total population found along the robot's path and the distance to the desired distribution of age.}
	\label{fig:pareto}
\end{figure}

\begin{figure}[t]
	\centering
	\def\tc{0} 
	\def\bc{0} 
	\def\lc{2.5} 
	\def\rc{2.5} 
	\includegraphics[width=0.32\linewidth, trim=\lc cm \bc cm \rc cm \tc cm, clip=true]{fig-new/martimlog-mu10/age-opt-JScity-real-JScity/map-box-map-age-20-24.png}
	\includegraphics[width=0.32\linewidth, trim=\lc cm \bc cm \rc cm \tc cm, clip=true]{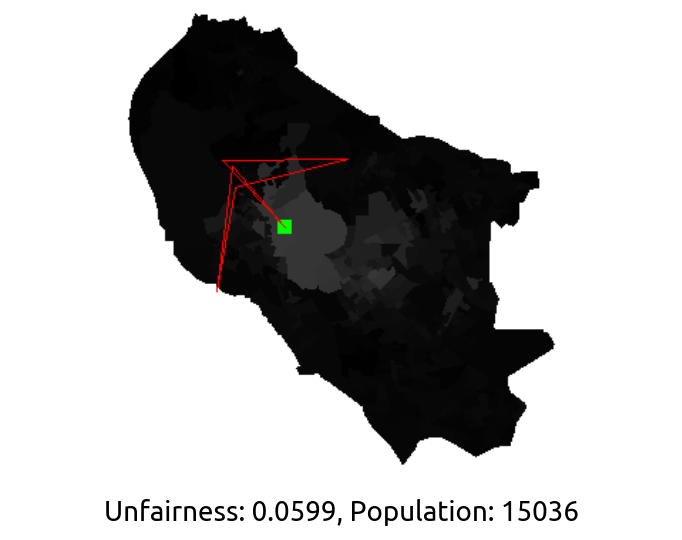}
	\includegraphics[width=0.32\linewidth, trim=\lc cm \bc cm \rc cm \tc cm, clip=true]{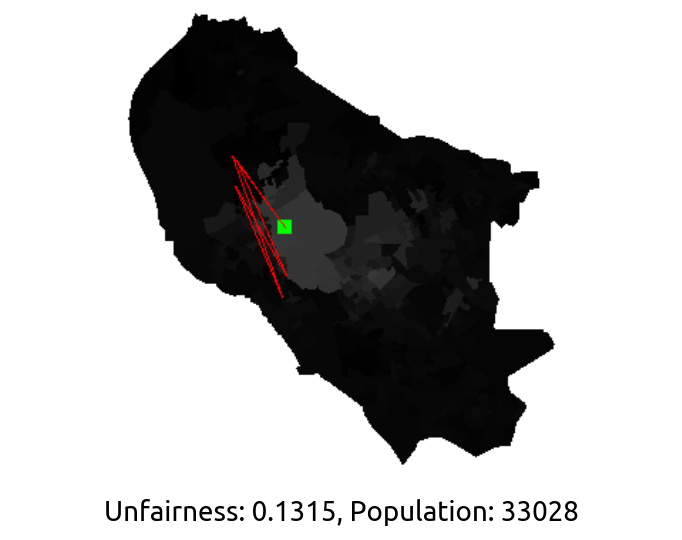} \\	
	\includegraphics[width=0.32\linewidth]{fig-new/martimlog-mu10/age-opt-JScity-real-JScity/age-box-length0400.png}
	\includegraphics[width=0.32\linewidth]{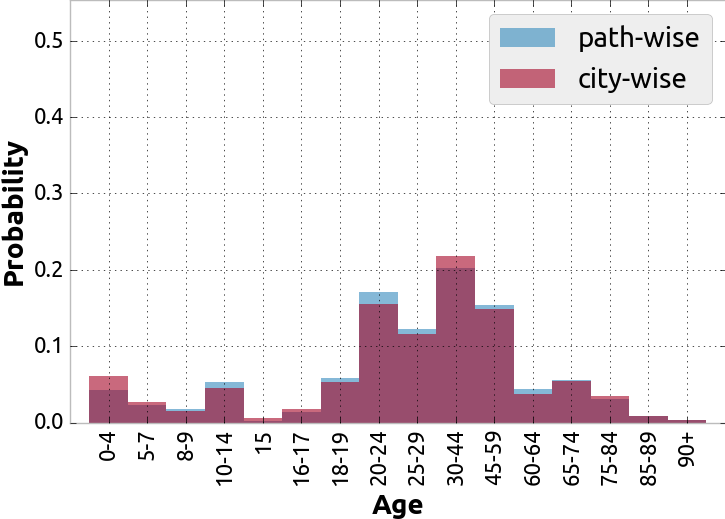}
	\includegraphics[width=0.32\linewidth]{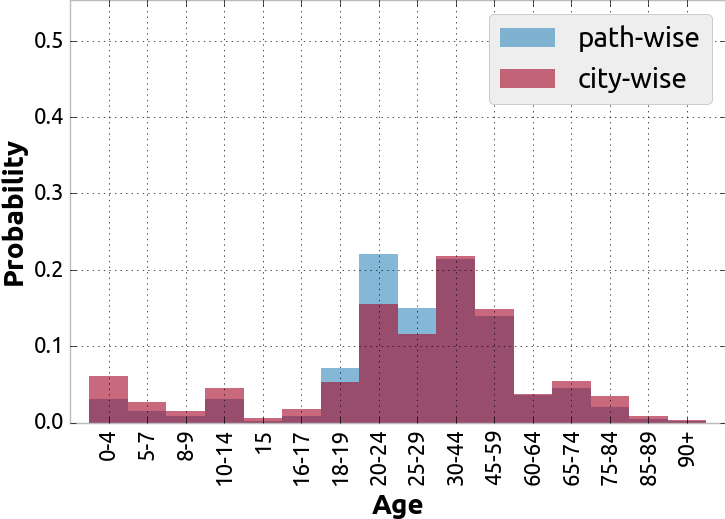}
	\caption{A thorough search path around the base (left) and the two extreme Pareto-optimal paths (middle and right).}
	\label{fig:paths}
\end{figure}

We found that to decrease unfairness it is necessary to reduce the total population found.
We arrived at this conclusion by computing the Pareto-front of two objectives: minimization of unfairness (Jensen-Shannon distance to perfect demographic parity for the sake of example) and maximization of efficiency (number of people found).
Figure~\ref{fig:pareto} shows this Pareto-front. Each point along the curve is a different path of different unfairness and efficiency values. Figure~\ref{fig:paths} shows the paths corresponding to the two extremes of the Pareto-front (i.e. lowest and highest unfairness). In our example, lowering unfairness requires lowering efficiency because older people also live more scattered and further away from the center than the younger population.
The figures also show that it was impossible for the method to find a path of strict fairness, i.e. where demographic parity is satisfied exactly, since the curve does not reach zero. This is understandable because, firstly, extreme luck should exist for a subsample of the city to exhibit exactly the same statistics as the city as a whole. Secondly, Pareto-front estimation methods such as the one we used are not guaranteed to find global minima since that would require exhaustively searching all possible paths within the map, which is unfeasible for the size of our problem.

In this situation, a stakeholder such as a decision-maker, emergency responder or policy maker could use the Pareto-fronts themselves to make a more informed decision about efficiency and fairness of the response, in a way that reflects the priorities and values at play in the specific situation. The decision-maker could select one of the solutions within the Pareto and thus explicitly weigh the conflicting objectives according to context.

\subsubsection{Current planning methods provide few guarantees}

The method we used to generate these paths is not a traditional one in robotics: it was based on evolutionary algorithms to compute Pareto-fronts \cite{Brandao2020aij}, which are algorithms that do not offer optimality guarantees. Traditional methods (e.g. A* with an admissible heuristic) could also have been used but they would require cost functions to be Markovian, i.e. the cost over a path to be equal to the sum of per-state costs. This is not the case for our protected characteristic fairness specification, and so a cumulative-cost approximation would have to be used.
In \cite{Brandao2020aij} we show that optimizing a cumulative-cost approximation would lead to twice higher unfairness in this case. 
So current methods provide few guarantees: they will either try to optimize the desired fairness metric without optimality guarantees, or they will optimize (with guarantees) a proxy metric that is different from what we care about.

\subsubsection{Fairness specifications can be counter-productive}

We found that some fairness specifications can lead to lower utility for all groups, compared to demographic parity. For example, we compared the result of promoting demographic parity on gender, with the result of promoting strict-equality affirmative action (i.e. a 50-50\% ratio). We found out that strict-equality can lead to lower utility for \emph{both} male and female classes. In other words, all classes were made worse-off just in order to find higher equality solutions.

Again, the use of fairness-related visualizations such as distributions and trade-offs is important to inform decision-makers about the impact and effectiveness of the technical choice. Furthermore, this example shows the advantage of simulating system deployments implementing different fairness definitions, in order to better evaluate and predict their results.
Finally, the example suggests that the process of specification---of selecting a definition of fairness---could be the product of an iterative design and validation process.

\subsubsection{Intuitive understanding}

The particular metric we have used for ``unfairness'' in Pareto-curves---distance to the fair distribution---also raises questions about which metrics are most intuitive for stakeholders to evaluate the degree to which a distributive principle is satisfied.
In this example we used Jensen-Shannon distance between distributions, but such a metric could arguably be considered unintuitive. The intuitiveness and interpretability of the choice of metric is an important topic of further research.

\subsubsection{Design is an iterative process}

A further issue that the example raises is that there are possibly multiple personal characteristics that a user or decision-maker cares about and would like to pay respect to. These may not be obvious from the onset of robot deployment.
For example, a disaster response team might program a fair navigation planner to respect a certain health and age-related feature, only to later find out they have a bias towards high-income neighborhoods that they would like to avoid. Alternatively, optimizing for fairness in one characteristic may introduce new biases in the paths that are again morally relevant. 

Part of the specification process will hence be in the discovery of which values matter for the application at hand, and how. This process can only succeed if there exist value-sensitive tools \cite{Friedman2013} that identify or support the identification of these issues. 
Our methodology in \cite{Brandao2020aij} was to develop a resource of applicable formal definitions that can be simulated to guide discussions and anticipate issues \emph{before} deployment, at the early stages of development.


\section{Discussion}
\label{sec:dissusion}


\subsection{Developing fair navigation planners}
\label{sec:dissusionDevelopingResponsibly}

As we have shown, building ``fair navigation planning'' algorithms certainly requires more than optimizing a fairness metric. 
We believe the ethical development and deployment of fair navigation algorithms requires work on multiple fronts.

\subsubsection{Transparency}

Planners should provide an intuitive understanding, through appropriate visualizations, statistics, or metrics, of the fairness characteristics of navigation plans.
Planners should be equipped with data and tools for the analysis of fairness across multiple variables and specifications, in order to allow impact and fairness-related issues to be spotted by stakeholders.

\subsubsection{Human autonomy}

In order to allow humans to responsibly use such tools, they need to be able to understand and control the trade-offs between fairness and other task objectives. This will require the use of visualization and human-in-the-loop design features. For example, Pareto-curves of the different objectives can serve as interesting visualizations and decision-aids. 
Additionally, users should be able to interface with the planning methods in order to include considerations of relevant expert knowledge, such as adding intermediate goals, biasing paths to certain solutions, or adjusting estimates of risk and utility (e.g. building damage or population density in the rescue case).

\subsubsection{Privacy}

Promoting distributive fairness over protected characteristics in navigation requires data on the distribution of these features themselves. This comes at the cost of having to gather such data, but also of potential privacy issues within the collection, analysis or security breaches of the data.

\subsubsection{Data availability and input}

Fairness with respect to locations cannot be fulfilled if there are missing locations in a map, and similarly for fairness on protected characteristics.
For fair navigation planning systems to be reliable, data and models of the relevant fairness features will be required. For example, population statistics over a map should be available to the planning algorithms. Some of these already exist---census data can be detailed in some countries---though they could arguably only be available in privileged circumstances. This calls for implementing ways for operators and stakeholders to populate maps with demographics from unstructured data and human knowledge.
Finally, even if demographic statistics are difficult to obtain in many humanitarian situations,  implementing fairness at the level of locations may still be desirable.


\subsection{Methodological contribution}
\label{sec:dissusionMethodologicalContribution}

Rather than only reflecting on general principles of fairness or requesting stakeholders to anticipate the consequences of an innovation, we believe that formal models and simulation-based investigation provide a more solid foundation on which to initiate discussions that can anticipate the consequences of an innovation. So, in the case of robot navigation it is possible to provide examples of particular decisions made in the design of an algorithm and their consequences. This can be a resource in stakeholder workshops where potential users, developers, policy makers and members of the general public seek to anticipate the implications of a technological innovation.  Modelling, formal specification and simulation can help provide a more systematic and informed foundation to such discussions, prior to any development taking place.


\section{Conclusion}
\label{sec:conclusion}

In this paper we explored the concept of fairness in the seemingly mundane, value-neutral, technical problem of robot navigation.
We showed that there is a fairness dimension to robot navigation, using a walkthrough example of a humanitarian rescue robot. We discussed how mobile robots will have to face similar dilemmas that humans already face.
We then applied theories of distributive justice to our navigation problem and used them to simulate robot deployment outcomes, ground discussions of fairness and design across multiple stakeholders, and anticipate issues.
We showed that fairness-aware navigation planners will involve efficiency-fairness trade-offs, that their design should be an iterative process of understanding the fairness issues of the context at hand, and that current planning methods have downsides that need be addressed.

This paper also sets the ground for a new research field of fair planning. Several challenges still lie ahead. Part of those are technical challenges of designing efficient, interpretable, formally verifiable and optimal methods. Another part is related to responsible innovation and value-sensitive design through appropriate analysis, visualization and participation tools.


\bibliographystyle{ACM-Reference-Format}
\bibliography{phil,allbib_robots_vision,allbib_brandao}


\begin{thebibliography}{13}


\ifx \showCODEN    \undefined \def \showCODEN     #1{\unskip}     \fi
\ifx \showDOI      \undefined \def \showDOI       #1{#1}\fi
\ifx \showISBNx    \undefined \def \showISBNx     #1{\unskip}     \fi
\ifx \showISBNxiii \undefined \def \showISBNxiii  #1{\unskip}     \fi
\ifx \showISSN     \undefined \def \showISSN      #1{\unskip}     \fi
\ifx \showLCCN     \undefined \def \showLCCN      #1{\unskip}     \fi
\ifx \shownote     \undefined \def \shownote      #1{#1}          \fi
\ifx \showarticletitle \undefined \def \showarticletitle #1{#1}   \fi
\ifx \showURL      \undefined \def \showURL       {\relax}        \fi
\providecommand\bibfield[2]{#2}
\providecommand\bibinfo[2]{#2}
\providecommand\natexlab[1]{#1}
\providecommand\showeprint[2][]{arXiv:#2}

\bibitem[\protect\citeauthoryear{Angwin, Larson, Mattu, and Kirchner}{Angwin
  et~al\mbox{.}}{2016}]%
        {Angwin2016}
\bibfield{author}{\bibinfo{person}{Julia Angwin}, \bibinfo{person}{Jeff
  Larson}, \bibinfo{person}{Surya Mattu}, {and} \bibinfo{person}{Lauren
  Kirchner}.} \bibinfo{year}{2016}\natexlab{}.
\newblock \bibinfo{title}{Machine Bias: there's software used across the
  country to predict future criminals. And it's biased against blacks.
  ProPublica 2016}.
\newblock
\newblock


\bibitem[\protect\citeauthoryear{Brandao}{Brandao}{2019}]%
        {Brandao2019fatecv}
\bibfield{author}{\bibinfo{person}{Martim Brandao}.}
  \bibinfo{year}{2019}\natexlab{}.
\newblock \showarticletitle{Age and gender bias in pedestrian detection
  algorithms}. In \bibinfo{booktitle}{\emph{Workshop on Fairness Accountability
  Transparency and Ethics in Computer Vision, CVPR}}.
\newblock
\urldef\tempurl%
\url{https://arxiv.org/abs/1906.10490}
\showURL{%
\tempurl}


\bibitem[\protect\citeauthoryear{Brandao, Jirotka, Webb, and Luff}{Brandao
  et~al\mbox{.}}{2020}]%
        {Brandao2020aij}
\bibfield{author}{\bibinfo{person}{Martim Brandao}, \bibinfo{person}{Marina
  Jirotka}, \bibinfo{person}{Helena Webb}, {and} \bibinfo{person}{Paul Luff}.}
  \bibinfo{year}{2020}\natexlab{}.
\newblock \showarticletitle{Fair navigation planning: a resource for
  characterizing and designing fairness in mobile robots}.
\newblock \bibinfo{journal}{\emph{Artificial Intelligence}}
  \bibinfo{volume}{282} (\bibinfo{year}{2020}).
\newblock
\showISSN{0004-3702}
\urldef\tempurl%
\url{https://doi.org/10.1016/j.artint.2020.103259}
\showDOI{\tempurl}


\bibitem[\protect\citeauthoryear{Buolamwini and Gebru}{Buolamwini and
  Gebru}{2018}]%
        {Buolamwini2018}
\bibfield{author}{\bibinfo{person}{Joy Buolamwini} {and}
  \bibinfo{person}{Timnit Gebru}.} \bibinfo{year}{2018}\natexlab{}.
\newblock \showarticletitle{Gender Shades: Intersectional Accuracy Disparities
  in Commercial Gender Classification}. In
  \bibinfo{booktitle}{\emph{Proceedings of the 1st Conference on Fairness,
  Accountability and Transparency}} \emph{(\bibinfo{series}{Proceedings of
  Machine Learning Research}, Vol.~\bibinfo{volume}{81})},
  \bibfield{editor}{\bibinfo{person}{Sorelle~A. Friedler} {and}
  \bibinfo{person}{Christo Wilson}} (Eds.). \bibinfo{publisher}{PMLR},
  \bibinfo{address}{New York, NY, USA}, \bibinfo{pages}{77--91}.
\newblock


\bibitem[\protect\citeauthoryear{Chouldechova}{Chouldechova}{2017}]%
        {Chouldechova2017}
\bibfield{author}{\bibinfo{person}{Alexandra Chouldechova}.}
  \bibinfo{year}{2017}\natexlab{}.
\newblock \showarticletitle{Fair Prediction with Disparate Impact: A Study of
  Bias in Recidivism Prediction Instruments}.
\newblock \bibinfo{journal}{\emph{Big Data}} \bibinfo{volume}{5},
  \bibinfo{number}{2} (\bibinfo{year}{2017}), \bibinfo{pages}{153--163}.
\newblock
\urldef\tempurl%
\url{https://doi.org/10.1089/big.2016.0047}
\showDOI{\tempurl}


\bibitem[\protect\citeauthoryear{Friedman, Kahn, Borning, and
  Huldtgren}{Friedman et~al\mbox{.}}{2013}]%
        {Friedman2013}
\bibfield{author}{\bibinfo{person}{Batya Friedman}, \bibinfo{person}{Peter~H
  Kahn}, \bibinfo{person}{Alan Borning}, {and} \bibinfo{person}{Alina
  Huldtgren}.} \bibinfo{year}{2013}\natexlab{}.
\newblock \showarticletitle{Value sensitive design and information systems}.
\newblock In \bibinfo{booktitle}{\emph{Early engagement and new technologies:
  Opening up the laboratory}}. \bibinfo{publisher}{Springer},
  \bibinfo{pages}{55--95}.
\newblock


\bibitem[\protect\citeauthoryear{Golub, Marcantonio, and Sanchez}{Golub
  et~al\mbox{.}}{2013}]%
        {Golub2013}
\bibfield{author}{\bibinfo{person}{Aaron Golub}, \bibinfo{person}{Richard~A
  Marcantonio}, {and} \bibinfo{person}{Thomas~W Sanchez}.}
  \bibinfo{year}{2013}\natexlab{}.
\newblock \showarticletitle{Race, space, and struggles for mobility:
  Transportation impacts on African Americans in Oakland and the East Bay}.
\newblock \bibinfo{journal}{\emph{Urban Geography}} \bibinfo{volume}{34},
  \bibinfo{number}{5} (\bibinfo{year}{2013}), \bibinfo{pages}{699--728}.
\newblock


\bibitem[\protect\citeauthoryear{Holzapfel, Sturm, and Coeckelbergh}{Holzapfel
  et~al\mbox{.}}{2018}]%
        {Holzapfel2018}
\bibfield{author}{\bibinfo{person}{Andre Holzapfel}, \bibinfo{person}{Bob
  Sturm}, {and} \bibinfo{person}{Mark Coeckelbergh}.}
  \bibinfo{year}{2018}\natexlab{}.
\newblock \showarticletitle{Ethical dimensions of music information retrieval
  technology}.
\newblock \bibinfo{journal}{\emph{Transactions of the International Society for
  Music Information Retrieval}} (\bibinfo{year}{2018}).
\newblock


\bibitem[\protect\citeauthoryear{Lin}{Lin}{2015}]%
        {Lin2015}
\bibfield{author}{\bibinfo{person}{Patrick Lin}.}
  \bibinfo{year}{2015}\natexlab{}.
\newblock \bibinfo{booktitle}{\emph{Why Ethics Matters for Autonomous Cars}}.
\newblock \bibinfo{publisher}{Springer Berlin Heidelberg},
  \bibinfo{address}{Berlin, Heidelberg}, \bibinfo{pages}{69--85}.
\newblock
\showISBNx{978-3-662-45854-9}
\urldef\tempurl%
\url{https://doi.org/10.1007/978-3-662-45854-9_4}
\showDOI{\tempurl}


\bibitem[\protect\citeauthoryear{Merin, Ash, Levy, Schwaber, and Kreiss}{Merin
  et~al\mbox{.}}{2010}]%
        {Merin2010}
\bibfield{author}{\bibinfo{person}{Ofer Merin}, \bibinfo{person}{Nachman Ash},
  \bibinfo{person}{Gad Levy}, \bibinfo{person}{Mitchell~J Schwaber}, {and}
  \bibinfo{person}{Yitshak Kreiss}.} \bibinfo{year}{2010}\natexlab{}.
\newblock \showarticletitle{The Israeli field hospital in Haiti-ethical
  dilemmas in early disaster response}.
\newblock \bibinfo{journal}{\emph{New England Journal of Medicine}}
  \bibinfo{volume}{362}, \bibinfo{number}{11} (\bibinfo{year}{2010}),
  \bibinfo{pages}{e38}.
\newblock


\bibitem[\protect\citeauthoryear{O'Math{\'u}na, Gordijn, and
  Clarke}{O'Math{\'u}na et~al\mbox{.}}{2013}]%
        {DisasterBioethics2013}
\bibfield{author}{\bibinfo{person}{D{\'o}nal~P O'Math{\'u}na},
  \bibinfo{person}{Bert Gordijn}, {and} \bibinfo{person}{Mike Clarke}.}
  \bibinfo{year}{2013}\natexlab{}.
\newblock \bibinfo{booktitle}{\emph{Disaster bioethics: Normative issues when
  nothing is normal}}. Vol.~\bibinfo{volume}{2}.
\newblock \bibinfo{publisher}{Springer Science \& Business Media}.
\newblock


\bibitem[\protect\citeauthoryear{Walker}{Walker}{2012}]%
        {Walker2012}
\bibfield{author}{\bibinfo{person}{Gordon Walker}.}
  \bibinfo{year}{2012}\natexlab{}.
\newblock \bibinfo{booktitle}{\emph{Environmental justice: concepts, evidence
  and politics}}.
\newblock \bibinfo{publisher}{Routledge}.
\newblock


\bibitem[\protect\citeauthoryear{Winfield and Jirotka}{Winfield and
  Jirotka}{2018}]%
        {Winfield2018}
\bibfield{author}{\bibinfo{person}{Alan F.~T. Winfield} {and}
  \bibinfo{person}{Marina Jirotka}.} \bibinfo{year}{2018}\natexlab{}.
\newblock \showarticletitle{Ethical governance is essential to building trust
  in robotics and artificial intelligence systems}.
\newblock \bibinfo{journal}{\emph{Philosophical Transactions of the Royal
  Society A: Mathematical, Physical and Engineering Sciences}}
  \bibinfo{volume}{376}, \bibinfo{number}{2133} (\bibinfo{year}{2018}),
  \bibinfo{pages}{20180085}.
\newblock
\urldef\tempurl%
\url{https://doi.org/10.1098/rsta.2018.0085}
\showDOI{\tempurl}


\end{thebibliography}


\end{document}